\documentclass[AMA,STIX1COL]{WileyNJD-v2}

\usepackage{amsmath,amsthm,hyperref}
\usepackage{color, soul} 
\usepackage{graphicx, caption}
\usepackage{mathtools}
\usepackage{comment}
\usepackage{xcolor}
\usepackage{listings} 
\usepackage{threeparttable} 

\definecolor{codegreen}{rgb}{0,0.6,0}
\definecolor{codegray}{rgb}{0.5,0.5,0.5}
\definecolor{codepurple}{rgb}{0.58,0,0.82}
\definecolor{backcolour}{rgb}{0.95,0.95,0.92}

\lstdefinestyle{mystyle}{
    backgroundcolor=\color{backcolour},
    commentstyle=\color{codegreen},
    keywordstyle=\color{blue},
    numberstyle=\tiny\color{codegray},
    stringstyle=\color{codepurple},
    basicstyle=\ttfamily\footnotesize,
    breakatwhitespace=false,
    breaklines=true,
    captionpos=b,
    keepspaces=true,
    numbers=left,
    numbersep=5pt,
    showspaces=false,
    showstringspaces=false,
    showtabs=false,
    tabsize=2,
    language = R,
    morekeywords = {inprod, ~, in},
    deletekeywords = {dim, sum, beta, gamma, t, q}
}

\articletype{Research Article}

\received{}
\revised{}
\accepted{}

\def\0{\mathbf 0}
\def\1{\mathbf 1}

\begin{document}

\title{A complex meta-regression model to identify effective features of interventions from multi-arm, multi-follow-up trials}

\author[1]{Annabel L Davies*}

\author[1]{Julian PT Higgins}

\authormark{A. DAVIES \textsc{et al}}

\address[1]{\orgdiv{Bristol Medical School}, \orgname{University of Bristol}, \orgaddress{\state{Bristol}, \country{United Kingdom}}}

\corres{*\email{annabel.davies@bristol.ac.uk}}

\presentaddress{Canynge Hall, Clifton, Bristol, BS8 2PN}

\abstract{Network meta-analysis (NMA) combines evidence from multiple trials to compare the effectiveness of a set of interventions. In public health research, interventions are often complex, made up of multiple components or features. This makes it difficult to define a common set of interventions on which to perform the analysis. One approach to this problem is component network meta-analysis (CNMA) which uses a meta-regression framework to define each intervention as a subset of components whose individual effects combine additively. In this paper, we are motivated by a systematic review of complex interventions to prevent obesity in children. Due to considerable heterogeneity across the trials, these interventions cannot be expressed as a subset of components but instead are coded against a framework of characteristic features. To analyse these data, we develop a bespoke CNMA-inspired model that allows us to identify the most important features of interventions. We define a meta-regression model with covariates on three levels: intervention, study, and follow-up time, as well as flexible interaction terms. By specifying different regression structures for trials with and without a control arm, we relax the assumption from previous CNMA models that a control arm is the absence of intervention components. Furthermore, we derive a correlation structure that accounts for trials with multiple intervention arms and multiple follow-up times. Although our model was developed for the specifics of the obesity data set, it has wider applicability to any set of complex interventions that can be coded according to a set of shared features.}

\keywords{Component network meta-analysis, meta-regression, complex interventions, multi-arm trials, multiple follow-up times}

\jnlcitation{\cname{%
\author{A. L. Davies}, and
\author{J. P. T. Higgins}} (\cyear{2023}),
\ctitle{A complex meta-regression model to identify effective features of interventions from multi-arm, multi-follow-up trials}, \cvol{2023;00:1--21}.}

\maketitle

\section{Introduction}

Network meta-analysis (NMA) is a technique, most commonly used in medical statistics, to combine evidence from multiple trials that target the same clinical condition.\cite{DIAS:2018} Compared with an individual trial, synthesizing evidence across multiple trials leads to more precise estimates of treatment performance and improves the generalizability of findings.\cite{CochraneBook} Conventional `pairwise' meta-analysis focuses on trials that compare the same two interventions. This produces a single statistic that summarizes the relative effect of one intervention compared with the other. NMA, on the other hand, is a more recent development that aims to make inferences about multiple competing treatments. \cite{Hig:White:1996, Lumley:2002} In particular, NMA makes use of `indirect evidence' to draw comparisons between interventions that have never been compared in a single trial. \cite{Bucher:1997} Indirect evidence refers to the idea that we can infer the relationship between two treatments, A and B, via their comparisons with some common third treatment C. By defining a network of intervention nodes connected by the trials in which they appear, we are able to combine direct and indirect evidence about all interventions in the network. \cite{Lu:Ades:2004, Caldwell:2005}

A complex intervention is one that consists of multiple, possibly interacting, components.\cite{Craig:2008, Lewin:2017}  In mental health research, for example, an intervention may combine multiple types of therapy (behavioural, cognitive and/or educational) or involve a combination of physiological and psychological strategies (e.g. drugs and counselling). Multi-component interventions are increasingly common in public health research,\cite{Petropoulou:2021} but their complexity makes them difficult to analyse using standard synthesis techniques. In the presence of complex interventions, the task of defining a set of common treatment nodes is not straightforward and can lead to disconnected networks that prohibit a coherent synthesis of the data. A common approach to this problem is to `lump'\cite{Welton:2009} interventions together based on some broad characteristic (e.g. `group therapy' vs `individual therapy') in order to connect the network. However, this is likely to introduce heterogeneity due to the variety of interventions included within each treatment category.

A common approach for dealing with heterogeneity in meta-analysis is meta-regression.\cite{TSD3} Similar to standard regression techniques, meta-regression involves predicting an outcome variable based on the values of one or more explanatory variables.\cite{CochraneBook, Thompson:2002} In this context, the outcome variable is the intervention effect and the explanatory variables are selected study characteristics. As with meta-analysis, study estimates are weighted by their precision and random effects terms can be included to account for any residual heterogeneity that is not captured by explanatory variables. Similar to NMA, network meta-regression is an extension of meta-regression to include multiple treatments. In fact, NMA itself can be framed as a meta-regression by modelling the different interventions as explanatory variables.\cite{Salanti:2008, Efthimiou:2016}

To account for multi-component interventions, Welton et al \cite{Welton:2009} used a meta-regression framework to define a `component network meta-analysis' (CNMA) model. In this model, each intervention is defined as a subset of individual components which are shared across the trials. The model estimates the effect of each component, assuming they combine additively. In other words, the effect of an intervention is set equal to the sum of the effects of its components. Welton et al also specified models that allow for interactions between the components.  In subsequent work, Freeman et al \cite{Freeman:2018} extended Welton et al's model to include additional dependence on study-level characteristics (e.g. baseline risk), while R{\"u}cker et al \cite{Rucker:2020} described the CNMA models in a frequentist framework. More recently, Konnyu et al \cite{Konnyu:2023} used a similar `response-surface' meta-regression approach to model associations of intervention components and other modifiers with post-treatment observations.

While the use of CNMA models is slowly increasing, they have not seen widespread uptake and there has been limited methodological development since Welton et al's seminal paper.\cite{Petropoulou:2021} In this article we refine and extend the existing CNMA methodology to analyse data from a systematic review of interventions to prevent obesity in children. The interventions included in the review include complex multi-component strategies. However, due to large heterogeneity across a variety of features, the task of defining a set of discrete components to characterize each intervention is not straightforward. In previous work, we instead developed an analytic framework to code each intervention according to a set of characteristic features believed to have an impact on obesity.\cite{Protocol}

To capture the varied and complex nature of the interventions, the analytic framework includes features related to context and implementation that cannot be interpreted as straightforward `components'. In contrast to other CNMA models, this means that a control arm is not defined by the absence of intervention components. To account for these complexities, we develop a bespoke meta-regression style model that aims to evaluate which intervention features are most strongly associated with effective obesity prevention. We model the data on the contrast level, defining intervention coefficients with respect to a control arm. We specify different models (with shared parameters) for trials with and without a control arm using the CNMA assumption that comparisons involving the same components (or in our case, the same features) cancel out.\cite{Rucker:2020} Therefore, the model for trials without a control arm only includes terms that represent differences between the interventions in that trial. Taking inspiration from existing CNMA models, we include study-level characteristics \cite{Freeman:2018} and allow for interactions between covariates.\cite{Welton:2009} We model multi-arm trials using standard NMA methodology and extend these methods to allow for observations at multiple follow-up times. We treat follow-up times as a categorical covariate and construct covariance structures to account for correlations between measurements made in the same trial. 

The model we develop extends existing CNMA models to facilitate analysis of interventions that cannot be broken down into `components' but can instead be characterized by a set of descriptive features. We allow for flexibility in terms of multi-arm trials, dependence on study characteristics, and interactions between covariates, and go beyond existing methodology to account also for multiple follow-up times. Therefore, as well as catering for the requirements of the obesity data set, the model  has applicability to the analysis of complex interventions more generally.

We begin by describing the motivating data set in Section \ref{sec:data}. In Section \ref{sec:notation} we set out our notation. This includes observations of our outcome variable, the three levels of covariate: intervention features, study characteristics and follow-up time, and interaction terms. We describe the development of our model in Section \ref{sec:Model}. We begin with the simplest case, assuming all trials are two-arm, have only one follow-up time and involve comparisons with a control arm. We then build up the model to include multi-arm trials and multi-follow-up trials. Finally, we use this to specify a model for trials without a control arm. Next, in Section \ref{sec:Application} we perform a preliminary analysis on the motivating data to demonstrate our model. We describe the specifics of the model set-up and its implementation in Section \ref{sec:Methods} and present the initial results in Section \ref{sec:Results}. Finally, we summarize and discuss our work in Section \ref{sec:Discuss} and outline our plans for subsequent detailed analysis of this data set.

\section{Motivating data set}\label{sec:data}
Our work is motivated by a systematic review of nearly 250 randomized controlled trials of interventions that aim to prevent obesity in children aged 5-18 years.\cite{Cochrane5to11, Cochrane12to18} The interventions vary significantly both within and across trials making it difficult to compare their effects using standard NMA methodology. In previous analyses of this data set,\cite{Cochrane5to11, Cochrane12to18} we `lumped' \cite{Welton:2009} the interventions together into whether they targeted diet, activity or a combination of both. While we observed some positive intervention effects relative to control, the results were largely heterogeneous. This is likely caused by the variation in the types of interventions within each category. For example, interventions coded as `diet' included strategies such as nutritional education lessons, changes to the school canteen menu, community cooking lessons, recipe cards for parents, and changes to local corner shops. Some interventions involved a combination of multiple strategies and we also observed large variability in the duration and intensity of their implementation. To capture these characteristics we developed an analytic framework to code the interventions according to a set of features believed to impact the intervention effect. We will describe details of the framework and how it was developed in a future paper; for an overview of this work see our protocol.\cite{Protocol}

To develop and test our model, we focus on trials which report measures of zBMI. zBMI is an age- and sex-standardized measure of body mass index (BMI) which is commonly used to define weight status in children.\cite{NICE} The trials in the data set report a mixture of arm-level and contrast-level data. Contrast-level measurements are mean differences in change from baseline between each intervention and a trial-specific reference arm. In our previous analysis of this data set we calculated contrast-level estimates from the arm-level data.\cite{Cochrane5to11, Cochrane12to18} Since the majority of the studies are cluster randomized controlled trials (CRCTs), where necessary, we adjusted the standard errors on our mean differences to account for clustering effects.\cite{CochraneBook, Cochrane5to11, Cochrane12to18} These mean differences and standard errors (either reported in the trial or calculated from arm-level measurements) make up the observations for our model. The data set involves both multi-arm trials (more than two interventions) and trials which report measurements at multiple follow-up times. Therefore, each trial may be associated with more than one contrast-level observation. 

\section{Notation}\label{sec:notation}
\subsection{Observations of relative treatment effects}
Each trial $i=1,\hdots, N$ has $A_i$ arms and  $T_i$ follow-up times and is associated with a $T_i(A_i-1)\times 1$ vector of observations, $\boldsymbol{y}_i = (\boldsymbol{y}_{i,1}, \boldsymbol{y}_{i,2}, \hdots, \boldsymbol{y}_{i,T_i})^\top =  (y_{i,1}^{(1)}, \hdots, y_{i,1}^{(A_i-1)}, \hdots, y_{i,T_i}^{(1)}, \hdots, y_{i,T_i}^{(A_i-1)})^\top$. Each $y_{i,t}^{(k)}$ is a mean difference in change from baseline at time $t$ between intervention $k$ and a trial-specific reference arm. In other words, they are observations of the trial-specific relative treatment effect of intervention $k$ in trial $i$ at follow-up time $t$. To simplify the notation we suppress the dependence on the baseline time, $b$, and the reference arm, $r$. In setting up our model we will occasionally use the more explicit notation,
\begin{align}
	y_{i,t}^{(k)} = y_{i,tb}^{(kr)},
\end{align}
where $k=1,\hdots, A_i-1$ and $t=1,\hdots,T_i$. In most trials, the reference arm is a control arm involving no active intervention, $r=C$. We refer to these trials as `control comparison studies'. Alternatively, trials without a control arm use an active intervention as the reference, $r=A_i$. We refer to these as `active comparison studies'.  

Each observation of mean difference is associated with a within-study variance (squared standard error), $v_{i,t}^{(k)}=\text{var}(y_{i,t}^{(k)})$, representing the precision of the measurement. We define the vector of variances in the same way as the observations, $\boldsymbol{v}_i = (\boldsymbol{v}_{i,1}, \boldsymbol{v}_{i,2}, \hdots, \boldsymbol{v}_{i,T_i})^\top =  (v_{i,1}^{(1)}, \hdots, v_{i,1}^{(A_i-1)}, \hdots, v_{i,T_i}^{(1)}, \hdots, v_{i,T_i}^{(A_i-1)})^\top$.

To develop our model, we will write the mean differences in terms of arm-level data. We define $d_{i,tb}^{(k)}$ as the mean change in zBMI from baseline to time point $t$ in arm $k$ of trial $i$ . For brevity, 
we will refer to this as a `change score'. We define $m_{i,t}^{(k)}$ as the mean zBMI in arm $k$ of trial $i$ at time $t$. Using this notation we can write
\begin{align} \label{eq:y_deconstruct}
		y_{i,t}^{(k)} &= d_{i,tb}^{(k)} - d_{i,tb}^{(r)} = (m_{i,t}^{(k)} - m_{i,b}^{(k)}) - (m_{i,t}^{(r)} - m_{i,b}^{(r)}),
\end{align}
where we have used $r$ to label the reference arm and $b$ to label the baseline time.

\subsection{Covariates}
The aim of our model is to identify characteristics of interventions that lead to the best outcomes. It is likely that such effects will also depend on other factors such as characteristics of the study and the length of follow-up. For example, intervention effects may vary depending on the country in which the trial was conducted. Similarly, certain interventions may cause significant changes in the short term but have limited long term impact. To capture these effects, we include covariates on three levels: intervention, study, and follow-up time. We introduce each of these in turn.

\subsubsection{Intervention-level covariates}
We assume that the analytic framework can be used to characterize interventions in terms of $n$ binary variables. Each intervention in each trial is then associated with a set of binary intervention covariates. Control arms, defined as `no active intervention', are not coded according to the analytic framework and therefore, have no associated intervention covariates. We label the intervention covariates as $x_{ij}^{(k)} \in \{0,1\}$, $i=1,\hdots,N$, $j=1,\hdots,n$ where $k=1,\hdots,A_i-1$ for control comparison studies and  $k=1,\hdots,A_i$ for active comparison studies. We assume the intervention features are fixed over time and therefore each $x_{ij}^{(k)}$ is independent of the follow-up time $t$.

\subsubsection{Study-level covariates}

We consider $p$ study-level covariates such that each trial $i$ is associated with $p$ variables, $z_{ij}$, $j=1,\hdots,p$. In a given trial, the study-level covariates are common to all treatment effects relative to the reference arm and do not vary with time. Therefore, they do not depend on the intervention $k$ or time point $t$.

\subsubsection{Follow-up time}
In our motivating data set follow-up times are categorized into short (12 weeks to $<$ 9 months), medium (9 to $<$ 15 months) and long term ($\geq$ 15 months). Each trial is associated with a maximum of three follow-up time observations, but most report only one or two. Based on the available data, we choose to model follow-up time as a categorical covariate. In general, we define $q$ as the number of follow-up time categories (e.g. $q=3$ in our data). For each study, we assume only one observation per time period such that the maximum number of follow-up times in study $i$ is $\max(T_i)=q$. 

A categorical variable with $q$ categories can be represented as $q-1$ dummy binary variables. Therefore, an observation at follow-up time $t$ in trial $i$ is associated with $q-1$ binary covariates, $w_{i,jt}\in\{0,1\}$, $j=1,\hdots,q-1$, which define the time period in which the observation falls. For example, to describe the $q=3$ follow-up time categories in the motivating data, we choose `short term' as our reference time point and define $q-1=2$ dummy variables $(w_{i,1t}, w_{i,2t})$. Here, $w_{i,1t}\in\{0,1\}$ indicates whether time $t$ in trial $i$ is medium term or not, and $w_{i,2t}\in\{0,1\}$ indicates whether it is long term. The three possible realizations of $(w_{i,1t}, w_{i,2t})$ correspond to the three follow-up time periods,
\begin{eqnarray}
	(w_{i,1t}, w_{i,2t})=(0,0) & \hspace{5pt}\xrightarrow{} & \text{short term}, \\
	(w_{i,1t}, w_{i,2t})=(1,0) & \hspace{5pt}\xrightarrow{} & \text{medium term},\\
	(w_{i,1t}, w_{i,2t})=(0,1) & \hspace{5pt}\xrightarrow{} & \text{long term}.
\end{eqnarray}
Note that a single follow-up time $t$ cannot be both medium and long term so the pair $(w_{i,1t}, w_{i,2t})=(1,1)$ is prohibited. We assume that every arm in a study is observed at the same follow-up times meaning the $w_{i,jt}$ terms are independent of the intervention arm $k$. 

\subsubsection{Interactions}
To improve the flexibility of our model, we will include interactions between covariates. Interactions allow for the effect of one covariate to vary depending on the value or level of another covariate. That is, the effect of two covariates combined could be larger or smaller than the sum of their individual effects. 

To include an interaction between two variables $x_1$ and $x_2$ in a meta-regression model, we simply add an additional term $+\eta x_1 x_2$ where $\eta$ is the parameter quantifying the extent of the interaction. In our model we will allow for interactions between and within the different levels of covariate. These could be pairwise interactions or interactions between more than two covariates. Let us assume that we wish to include $l$ interactions. We write $\mathcal{J}_{i,jt}^{(k)}$ for the $j^{th}$ set of interaction covariates for intervention $k$ at follow-up time $t$ in trial $i$. Since we allow for interactions between any combination of covariates on any level, the interactions may depend on the study, intervention, and follow-up time. Each  $\mathcal{J}_{i,jt}^{(k)}$ is a product between a set of covariates $\in \{x_{i,j'}^{(k)}, z_{i,j''}, w_{i,j'''t}; j'=1,\hdots,n, j''=1,\hdots,p, j'''=1,\hdots,q-1\}$. For example, we might choose $l=3$ and define
\begin{align}
	\mathcal{J}_{i,1t}^{(k)} &= z_{i,1}z_{i,3},\\
	\mathcal{J}_{i,2t}^{(k)} &= x_{i,2}^{(k)} w_{i,2t}, \\
	\mathcal{J}_{i,3t}^{(k)} &= x_{i,1}^{(k)} x_{i,2}^{(k)} z_{i,2}.
\end{align}
Here, the first interaction term is between two study-level covariates and depends only on the study $i$. The second interaction is between an intervention feature and the second follow-up time (e.g. long term) and therefore depends on study, intervention and follow-up time. The third is a three-way interaction between two intervention features and a study characteristic and depends on the study and intervention arm. 

\section{Model}\label{sec:Model}
In the following we describe the development of our model. We begin with the simplest case, assuming every trial is a control comparison study with two arms and one follow-up time. We then extend this to include (i) multi-arm trials and (ii) multi-follow-up trials. Finally, we use this model to define a model for active comparison studies. 

\subsection{Control comparison studies}
\subsubsection{Two-arm, single follow-up trials}
 
For two-arm trials with one follow-up time, the vectors $\boldsymbol{y}_i$ and $\boldsymbol{v}_i$ reduce to the scalars $y_i$ and $v_i$ representing the observed mean difference (relative to control, $C$) and its variance in trial $i$. In other words, we can suppress the indices $k$ and $t$. We assume the observations $y_i$ are normally distributed about some trial-specific relative effect, $\delta_{i}$,
\begin{align}
	y_i &\sim N(\delta_{i}, v_i).
\end{align}
To allow the intervention effects to vary between trials, we model $\delta_i$ as a random variable centred on a trial-specific fixed effect, $\theta_i$,
\begin{align}
	\delta_i &\sim N(\theta_i, \tau^2),
\end{align}
where the variance $\tau^2$ is the between-trial heterogeneity. For the fixed effect, we define a meta-regression style model assuming that the effects of all covariates are additive and allowing for $l$ interaction terms,
\begin{align}
	\theta_i &= \alpha + \sum_{j=1}^{n} \beta_j x_{i,j} + \sum_{j=1}^{p} \gamma_j z_{i,j} + \sum_{j=1}^{q-1} \phi_j w_{i,j} + \sum_{j=1}^{l} \eta_j\mathcal{J}_{i,j}.
\end{align}
Here, the intercept $\alpha$ is the effect, relative to control, of an intervention whose covariates are all set to 0. By including this intercept we are not simply defining a control arm as an intervention with all covariates equal to zero. For example, intervention features might represent intensity (high = 1, low = 0), duration (long = 1, short = 0) and ``fun factor" (fun = 1, boring = 0). The $\alpha$ parameter then allows for a low intensity, short, boring intervention to have a different effect compared with no intervention at all. 

The parameters $\boldsymbol{\beta}=(\beta_1, \hdots, \beta_n)^\top$, $\boldsymbol{\gamma}=(\gamma_1, \hdots, \gamma_p)^\top$, and $\boldsymbol{\phi}=(\phi_1, \hdots, \phi_{q-1})^\top$ are the regression coefficients for the intervention features, study-level covariates and follow-up time indicators respectively. Although it is possible to collect all the covariates into a vector and define a single set of regression coefficients, we choose to write the different covariate levels explicitly. The parameters $\boldsymbol{\eta}=(\eta_1, \hdots, \eta_l)^\top$ are the regression coefficients for the interaction parameters.

\subsubsection{Multi-arm, single follow-up trials} \label{sec:MultiArm-SingleFU}
We use standard NMA methodology \cite{Hig:White:1996, Franchini:2012} to extend our model to include multi-arm trials. Trials with $A_i$ arms contribute $A_i-1$ observations corresponding to the mean difference between each intervention $k=1,\hdots,A_i-1$ and the reference arm, $r=C$. Therefore, the vectors $\boldsymbol{y}_i$ and $\boldsymbol{v}_i$ now have dimensions $(A_i-1)\times 1$. Defining the vectors $\boldsymbol{\delta}_i = (\delta_i^{(1)}, \hdots, \delta_i^{(A_i-1)})^\top$ and $\boldsymbol{\theta}_i = (\theta_i^{(1)}, \hdots, \theta_i^{(A_i-1)})^\top$, we write our model as
\begin{align}\label{eq:y-del-multi-arm}
	\boldsymbol{y}_i &\sim N(\boldsymbol{\delta}_i, \boldsymbol{V}_i) \nonumber\\
	\boldsymbol{\delta}_i &\sim N(\boldsymbol{\theta}_i, \boldsymbol{\Sigma}_i),
\end{align}
where $\boldsymbol{V}_i$ and $\boldsymbol{\Sigma}_i$ are  $(A_i-1)\times(A_i-1)$ covariance matrices (defined further below). For each non-control arm $k$ in trial $i$, we model the fixed effects in the same way as the two-arm case, 
\begin{align}
	\theta_{i}^{(k)} &= \alpha + \sum_{j=1}^{n} \beta_j x_{i,j}^{(k)} + \sum_{j=1}^{p} \gamma_j z_{i,j} + \sum_{j=1}^{q-1} \phi_j w_{i,j} + \sum_{j=1}^{l} \eta_j\mathcal{J}_{i,j}^{(k)}, \label{eq:model-multi-arm}
\end{align}
where the intervention features and interactions depend on the intervention arm. The intercept, study-level covariates and follow-up time are independent of the intervention and hence have no index $k$. 

We capture correlations between observations of different interventions in the same trial via the covariance matrices in Equation (\ref{eq:y-del-multi-arm}). The matrix $\boldsymbol{V}_i$ models the random error in the observed relative effects due to sampling within the trial arms, 
\begin{align}
	 \boldsymbol{V}_i = \begin{pmatrix}
		v_i^{(1)} & \text{cov}(y_i^{(1)},y_i^{(2)}| \boldsymbol{\delta}_i) & \hdots & \text{cov}(y_i^{(1)},y_i^{(A_i-1)}| \boldsymbol{\delta}_i)\\
		\text{cov}(y_i^{(2)},y_i^{(1)}| \boldsymbol{\delta}_i) & v_i^{(2)} & \hdots & \text{cov}(y_i^{(2)},y_i^{(A_i-1)}| \boldsymbol{\delta}_i)\\
		\vdots & \vdots & \ddots & \vdots \\
		\text{cov}(y_i^{(A_i-1)},y_i^{(1)}| \boldsymbol{\delta}_i) & \text{cov}(y_i^{(A_i-1)},y_i^{(2)}| \boldsymbol{\delta}_i) & \hdots & v_i^{(A_i-1)}
	\end{pmatrix}. \label{eq:Err_multi1}
\end{align}
The diagonal elements are equal to the sampling variance of each $y_i^{(k)}$ and the off-diagonal elements are the covariances between each pair of measurements due to the within-study sampling. Using the definition of mean difference in Equation (\ref{eq:y_deconstruct}), we find that the covariance between observations in different arms of the same trial is equal to the variance of the change score in the reference arm, 
\begin{align}
	\text{cov}(y_i^{(k)},y_i^{(k')}| \boldsymbol{\delta}_i) &= \text{cov}(d_{i}^{(k)} - d_{i}^{(r)}, d_{i}^{(k')} - d_{i}^{(r)}) \\
	&= \text{cov}(d_{i}^{(k)}, d_{i}^{(k')}) - \text{cov}(d_{i}^{(k)}, d_{i}^{(r)}) - \text{cov}(d_{i}^{(r)}, d_{i}^{(k')}) + \text{cov}(d_{i}^{(r)}, d_{i}^{(r)})\\
	&=\text{var}(d_{i}^{(r)}),  \label{eq:var-ref}
\end{align}
where we have used the fact that the change from baseline measurements $d_i^{(k)}$ in each arm are independent.

The covariance matrix $\boldsymbol{\Sigma}_i$ models the trial-to-trial variation of the relative treatment effects. It describes the variance of the relative effects $\delta_{i}^{(k)}$, $k=1,\hdots,A_i-1$, and their correlations. We define it as
\begin{align}\label{eq:RE_multi1}
	\boldsymbol{\Sigma}_i = \begin{pmatrix}
		\tau^2 & \tau^2/2 & \hdots  \\
		\tau^2/2 & \tau^2 & \hdots\\
		\vdots & \vdots & \ddots
	\end{pmatrix},
\end{align}
where we have used the standard assumption from NMA that all trial-specific relative effects have the same variance, $\tau^2$, and that the correlation between any two relative effects in the same trial is $1/2$.\cite{Hig:White:1996} Based on the transitivity assumption, this ensures that the variance of the relative effect between any two interventions within a trial is also $\tau^2$. For details, refer to Appendix \ref{App:RE-arm}.

\subsubsection{Multi-arm, multi-follow-up trials}\label{sec:MultiArm-MultiFU}

We now extend our model to include trials which report observations at multiple follow-up times, taking our approach beyond standard NMA methodology. Based on the minimal longitudinal data available in our motivating data set we choose to model follow-up time as a categorical covariate. In doing so we follow a similar approach to Musekiwa et al (2016) \cite{Musekiwa:2016} who define a meta-regression model for a pairwise meta-analysis with observations at multiple time points. Our approach is similar to their `Model 5' however, the extra complexities in our model (multi-arm trials and component-style intervention features) lead us to make stronger assumptions about between-trial heterogeneity. We explain this in more detail below.

Each trial now reports $T_i(A_i-1)$ observations, where we recall that $T_i$ labels the number of follow-up times in trial $i$. The vectors, $\boldsymbol{y}_i$ and $\boldsymbol{v_i}$, take their full form with dimensions $T_i(A_i-1)\times 1$. Following from our previous models, we write
\begin{align} \label{eq:model-multi-arm-FU}
	&\boldsymbol{y}_i \sim N(\boldsymbol{\delta}_i, \boldsymbol{V}_i) \nonumber\\
	&\boldsymbol{\delta}_i \sim N(\boldsymbol{\theta}_i, \boldsymbol{\Sigma}_i) \nonumber\\
	&\theta_{i,t}^{(k)} = \alpha + \sum_{j=1}^{n} \beta_j x_{i,j}^{(k)} + \sum_{j=1}^{p} \gamma_j z_{i,j} + \sum_{j=1}^{q-1} \phi_j w_{i,jt} + \sum_{j=1}^{l} \eta_j\mathcal{J}_{i,jt}^{(k)},
\end{align}
where the vector $\boldsymbol{\delta}_i$ contains the effects $\delta_{i,t}^{(k)}$ of each intervention, $k=1,\hdots,A_i-1$, relative to control, $C$, at each follow-up time $t=1,\hdots,T_i$. Similarly, the vector $\boldsymbol{\theta}_i$ contains the fixed effect terms $\theta_{i,t}^{(k)}$. In the expression for the fixed effects, the intercept, intervention features and study-level covariates are independent of the follow-up time and therefore have no index $t$. 

As well as correlations due to multi-arm trials, we need to account for correlations due to multiple follow-up times. For the random effects, we essentially treat multi-follow-up trials in the same way as multi-arm trials. In multi-arm studies, different interventions $k$ have different $\{x_{i,j}^{(k)}\}$ covariates. In multi-follow-up studies, different time points $t$ have different $\{w_{i,jt}\}$ covariates.  To define the between-trial covariance matrix $\boldsymbol{\Sigma}_i$, we extend the standard NMA assumption of common between-trial heterogeneity to include multiple follow-up times. That is, we assume a common between-trial variance of $\tau^2$ for all arms and all time points. As before, this implies that the correlation between any pair of relative effects $(\delta_{i,t}^{(k)}, \delta_{i,t'}^{(k')})$ in a trial is $1/2$.  We provide details of this assumption in Appendix \ref{App:RE-FU}. The between-trial covariance matrix then takes the same form as Equation (\ref{eq:RE_multi1}), but now with larger dimensions $T_i(A_i-1)\times T_i(A_i-1)$.

The covariance matrix $\boldsymbol{V}_i$ captures correlations between outcome measurements within trials which have multiple interventions and multiple follow-up times. It has dimensions $T_i(A_i-1)\times T_i(A_i-1)$ and takes the form
\begin{align}\label{eq:Cov_mat}
	\boldsymbol{V}_i = \begin{pmatrix}
		\boldsymbol{V}_{i,1} & \boldsymbol{V}_{i,12} & \hdots & \boldsymbol{V}_{i,1T_i}\\
		\boldsymbol{V}_{i,21} & \boldsymbol{V}_{i,2} & \hdots & \boldsymbol{V}_{i,2T_i}\\
		\vdots & \vdots & \ddots & \vdots \\
		\boldsymbol{V}_{i,T_i1} & \boldsymbol{V}_{i,T_i2} & \hdots & \boldsymbol{V}_{i,T_i}
	\end{pmatrix},
\end{align}
where each diagonal element $\boldsymbol{V}_{i,t}$ is an $(A_i-1)\times(A_i-1)$ matrix representing the correlations between different arms at fixed time point $t$,
\begin{align}\label{eq:Vit}
	\boldsymbol{V}_{i,t} = \begin{pmatrix}
		v_{it}^{(1)} & \text{cov}(y_{it}^{(1)}, y_{it}^{(2)}| \boldsymbol{\delta}_i) & \hdots & \text{cov}(y_{it}^{(1)}, y_{it}^{(A_i-1)}| \boldsymbol{\delta}_i)\\
		\text{cov}(y_{it}^{(2)}, y_{it}^{(1)}| \boldsymbol{\delta}_i) & v_{it}^{(2)} & \hdots & \text{cov}(y_{it}^{(2)}, y_{it}^{(A_i-1)}| \boldsymbol{\delta}_i)\\
		\vdots & \vdots & \ddots & \vdots \\
		\text{cov}(y_{it}^{(A_i-1)}, y_{it}^{(1)}| \boldsymbol{\delta}_i) & \text{cov}(y_{it}^{(A_i-1)}, y_{it}^{(2)}| \boldsymbol{\delta}_i) & \hdots & v_{it}^{(A_i-1)}
	\end{pmatrix}.
\end{align}
In other words, each diagonal matrix in Equation (\ref{eq:Cov_mat}) is equivalent to the covariance matrix in Equation (\ref{eq:Err_multi1}) for multi-arm, single follow-up trials. The off-diagonal elements $\boldsymbol{V}_{i,tt'}$ are $(A_i-1)\times(A_i-1)$ matrices representing the covariances between time points $t$ and $t'$ and between different interventions,
\begin{align}\label{eq:Vittprime}
	\boldsymbol{V}_{i,tt'} = \begin{pmatrix}
		\text{cov}(y_{it}^{(1)}, y_{it'}^{(1)}| \boldsymbol{\delta}_i) & \text{cov}(y_{it}^{(1)}, y_{it'}^{(2)}| \boldsymbol{\delta}_i) & \hdots & \text{cov}(y_{it}^{(1)}, y_{it'}^{(n_i-1)}| \boldsymbol{\delta}_i)\\
		\text{cov}(y_{it}^{(2)}, y_{it'}^{(1)}| \boldsymbol{\delta}_i) & \text{cov}(y_{it}^{(2)}, y_{it'}^{(2)}| \boldsymbol{\delta}_i) & \hdots & \text{cov}(y_{it}^{(2)}, y_{it'}^{(n_i-1)}| \boldsymbol{\delta}_i)\\
		\vdots & \vdots & \ddots & \vdots \\
		\text{cov}(y_{it}^{(n_i-1)}, y_{it'}^{(1)}| \boldsymbol{\delta}_i) & 	\text{cov}(y_{it}^{(n_i-1)}, y_{it'}^{(2)}| \boldsymbol{\delta}_i) & \hdots & \text{cov}(y_{it}^{(n_i-1)}, y_{it'}^{(n_i-1)}| \boldsymbol{\delta}_i)\\
	\end{pmatrix}.
\end{align}
The structure of $\boldsymbol{V}_i$ is such that it captures within-trial correlations between observations of the following instances:
\begin{enumerate}
	\item $t=t'$, $k=k'$: the same time point and same intervention (diagonal elements of $\boldsymbol{V}_{i,t}$),
	\item $t=t'$, $k\neq k'$: the same time point and different interventions (off-diagonal elements of $\boldsymbol{V}_{i,t}$),
	\item $t\neq t'$, $k = k'$ different time points and the same intervention (diagonal elements of $\boldsymbol{V}_{i,tt'}$), and
	\item $t\neq t'$, $k\neq k'$ different time points and different interventions (off-diagonal elements of $\boldsymbol{V}_{i,tt'}$).
\end{enumerate}
In Appendix \ref{App:Covs} we evaluate each of these covariances in terms of the data. In doing so, we define the correlation coefficients $\rho_{y,tt'}$ and $\rho_{d,tt'}$. The former is the correlation between observations of mean difference for a given intervention (relative to the reference arm) at time points $t$ and $t'$, assumed to be common to all arms. The latter is the correlation between observations of the change score in the reference arm between $t$ and $t'$. Both coefficients are unknown but can be imputed from the data. For details, refer to the appendix.

\subsection{Active comparison studies}
For active comparison studies, the mean differences $y_{i,t}^{(k)}$ are not defined relative to a control arm, $C$. Instead, the reference arm is an active intervention. Without loss of generality, we choose this reference arm to be $r = A_i$. The study reports $A_i-1$ mean differences relative to this arm, defined as 
\begin{align}\label{eq:y-active}
	y_{i,t}^{(k)} = y_{i,tb}^{(kA_i)} = d_{i,tb}^{(k)} - d_{i,tb}^{(A_i)},
\end{align}
for $k=1,\hdots,A_i-1$. We treat each $A_i$-armed active comparison study as if it were an $(A_i+1)$-armed control comparison study where the measurement of the control arm is missing. If the control arm measurement was not missing, the equivalent control comparison study would report measurements
\begin{align}
	y_{i,t}^{(k)} = y_{i,tb}^{(kC)} = d_{i,tb}^{(k)} - d_{i,tb}^{(C)},
\end{align}
for $k=1,\hdots,A_i$. Inserting $-d_{i,tb}^{(C)}+d_{i,tb}^{(C)}$ into Equation (\ref{eq:y-active}) it is easy to show that measurements within a trial are transitive
\begin{align}\label{eq:y-trans}
	y_{i,tb}^{(kA_i)} = (d_{i,tb}^{(k)}- d_{i,tb}^{(C)}) - (d_{i,tb}^{(A_i)} - d_{i,tb}^{(C)}) = y_{i,tb}^{(kC)} - y_{i,tb}^{(A_iC)}.
\end{align}
Therefore, we define the active comparison model by taking the difference between two control comparison models in the same trial but with different interventions. That is, we insert Equation (\ref{eq:model-multi-arm-FU}) in for each $y_{i,tb}^{(kC)}$ in Equation (\ref{eq:y-trans}). This yields similar transitivity relations for the random and fixed trial-specific relative effects,
\begin{align}
 \delta_{i,tb}^{(kA_i)} &= \delta_{i,tb}^{(kC)} - \delta_{i,tb}^{(A_iC)}, \label{eq:delta_trans}\\
 \theta_{i,tb}^{(kA_i)} &= \theta_{i,tb}^{(kC)} - \theta_{i,tb}^{(A_iC)}. \label{eq:theta-trans}
\end{align}
We collect the $\delta_{i,tb}^{(kA_i)}$ and $\theta_{i,tb}^{(kA_i)}$ ($k=1,\hdots,A_i-1$, $t=1,\hdots, T_i$) terms in the vectors $\boldsymbol{\delta}_i$ and $\boldsymbol{\theta}_i$ respectively. The model for active comparison trials can then be written as
\begin{align} \label{eq:active-full}
	&\boldsymbol{y}_i \sim N(\boldsymbol{\delta}_i, \boldsymbol{V}_i) \nonumber\\
	&\boldsymbol{\delta}_i \sim N(\boldsymbol{\theta}_i, \boldsymbol{\Sigma}_i) \nonumber\\
	&\theta_{i,tb}^{(kA_i)} = \sum_{j=1}^{n} \beta_j (x_{i,j}^{(k)}-x_{i,j}^{(A_i)}) + \sum_{j=1}^{l} \eta_j(\mathcal{J}_{i,jt}^{(k)} - \mathcal{J}_{i,jt}^{(A_i)}),
\end{align}
where all terms in the regression model that do not depend on the intervention cancel out in the subtraction. Therefore, active comparison trials only contribute information about parameters which are associated with covariates that differ between the intervention $k$ and the reference arm $r=A_i$. In particular, this means that the active comparison regression model does not involve the intercept, study-level characteristics or follow-up time covariates. 

In defining the first two lines of Equation (\ref{eq:active-full}) we have used the fact that taking the difference between two normal distributions yields a normal distribution. In Appendix \ref{App:Cov_active} we show that both the covariance matrices $\boldsymbol{\Sigma}_i$ and $\boldsymbol{V}_i$ follow the same distributions as in the control comparison model, Equations (\ref{eq:RE_multi1}) and (\ref{eq:Cov_mat}) respectively.

\subsection{Model summary}
In summary, each trial supplies a $T_i(A_i-1)\times 1$ vector of observations $\boldsymbol{y}_i = (\boldsymbol{y}_{i,1}, \boldsymbol{y}_{i,2}, \hdots, \boldsymbol{y}_{i,T_i})^\top =  (y_{i,1}^{(1)}, \hdots, y_{i,1}^{(A_i-1)}, \hdots, y_{i,T_i}^{(1)}, \hdots, y_{i,T_i}^{(A_i-1)})^\top$. Each $y_{i,t}^{(k)}$ is a mean difference in change from baseline to time $t$ between arm $k$ and some trial-specific reference arm $r$. The reference arm is either a control arm, $r=C$, or another active intervention, $r=A_i$. We assume the observations follow a normal distribution centred on the trial-specific relative treatment effects, $\boldsymbol{\delta_i}$. The covariance matrix $\boldsymbol{V}_i$ captures correlations between measurements made in the same trial (due to multiple arms and/or follow-up times) and is assumed to be known. We assume a random effects model for the trial-specific effects, normally distributed about the fixed effects, $\boldsymbol{\theta}_i$. We model correlations between the random effects via the covariance matrix $\boldsymbol{\Sigma}_i$. This matrix depends on the heterogeneity variance, $\tau^2$, which we assume captures the between-trial variance for all arms and all follow-up times. The vectors $\boldsymbol{\delta_i}$ and $\boldsymbol{\theta}_i$ have the same dimensions and structure as the vector of observations $\boldsymbol{y}_i$. For the fixed effects, $\{\theta_{i,t}^{(k)}; k=1,\hdots, A_i-1, t=1,\hdots,T_i\}$,  we define a meta-regression model in terms of three levels of covariate (intervention, study and follow-up time) and allow for interactions between them. The form of the regression model depends on the reference arm of the trial. The full model is summarized as follows,
\begin{align} \label{eq:model-summary}
	&\boldsymbol{y}_i \sim N(\boldsymbol{\delta}_i, \boldsymbol{V}_i) \nonumber\\
	&\boldsymbol{\delta}_i \sim N(\boldsymbol{\theta}_i, \boldsymbol{\Sigma}_i) \nonumber\\[5pt]
	&\theta_{i,t}^{(k)} = \begin{cases}
		\alpha + \sum_{j=1}^{n} \beta_j x_{i,j}^{(k)} + \sum_{j=1}^{p} \gamma_j z_{i,j} + \sum_{j=1}^{q-1} \phi_j w_{i,jt} + \sum_{j=1}^{l} \eta_j\mathcal{J}_{i,jt}^{(k)} & \text{ for } r = C\\		
		\sum_{j=1}^{n} \beta_j (x_{i,j}^{(k)}-x_{i,j}^{(r)}) + \sum_{j=1}^{l} \eta_j(\mathcal{J}_{i,jt}^{(k)} - \mathcal{J}_{i,jt}^{(r)}) & \text{ for } r=A_i 
	\end{cases} \nonumber\\[8pt]
	&\boldsymbol{V}_i = \begin{pmatrix}
		\boldsymbol{V}_{i,1} & \boldsymbol{V}_{i,12} & \hdots & \boldsymbol{V}_{i,1T_i}\\
		\boldsymbol{V}_{i,21} & \boldsymbol{V}_{i,2} & \hdots & \boldsymbol{V}_{i,2T_i}\\
		\vdots & \vdots & \ddots & \vdots \\
		\boldsymbol{V}_{i,T_i1} & \boldsymbol{V}_{i,T_i2} & \hdots & \boldsymbol{V}_{i,T_i}
	\end{pmatrix} \hspace{10pt}
	\boldsymbol{\Sigma}_i = \begin{pmatrix}
		\tau^2 & \tau^2/2 & \hdots & \tau^2/2 \\
		\tau^2/2 & \tau^2 & \hdots & \tau^2/2\\
		\vdots & \vdots & \ddots & \vdots \\
		\tau^2/2 & \tau^2/2 & \hdots & \tau^2
	\end{pmatrix},
\end{align}
where matrices $\boldsymbol{V}_{i,t}$ and $\boldsymbol{V}_{i,tt'}$ are defined in Equations (\ref{eq:Vit}) and (\ref{eq:Vittprime}) respectively.

\section{Application}\label{sec:Application}

\subsection{Methods}\label{sec:Methods}
To demonstrate the application of our model, we perform a preliminary analysis of the motivating data set described in Section \ref{sec:data}. Based on our previous analysis of this data,\cite{Cochrane5to11, Cochrane12to18} we choose a small subset of covariates to investigate.

\subsubsection{Covariate definitions}
\textit{Intervention level covariates:} In our previous analyses,\cite{Cochrane5to11, Cochrane12to18} we were interested in the comparison of diet, activity and the combination of diet and activity. Our inclusion criteria require that each active intervention targets at least one of diet and activity meaning the combination (0,0) for these variables is not possible. Therefore, including diet, activity and an interaction term would cause the model to be over-parametrized. To avoid this, we code the three scenarios as two covariates, ``activity only" ($j=1$) and ``diet and activity" ($j=2$), where ``diet only" is treated as the reference. We also include the intervention factors ``intensity" ($j=3$) and ``duration" ($j=4$). Intensity is defined as `high' if the recipient engages with the intervention at least once a week and `low' otherwise. We dichotomize duration as `long' and `short' based on the median duration of all interventions in the data set. The intervention level covariates are then defined as follows,
\begin{align}
	(x_{i,1}^{(k)}, x_{i,2}^{(k)}) &= \begin{cases}
		(0,0) & \text{if intervention $k$ targets diet only}\\
		(1,0) & \text{if intervention $k$ targets activity only}\\
		(0,1) & \text{if intervention $k$ targets both diet and activity}.
	\end{cases}\nonumber\\
	x_{i,3}^{(k)} &= \begin{cases}
	1 & \text{if intervention $k$ is `high' intensity}\\
	0 & \text{if intervention $k$ is `low' intensity},
	\end{cases}\nonumber\\
	x_{i,4}^{(k)} &= \begin{cases}
	1 & \text{if intervention $k$ has a `long' duration}\\
	0 & \text{if intervention $k$ has a `short' duration}.
\end{cases}
\end{align}

\textit{study-level covariates:} Previously,\cite{Cochrane5to11, Cochrane12to18} we performed separate syntheses for trials that targeted different age groups (5-11 years or 12-18 years).  Therefore we choose ``age" as our one study-level covariate,
\begin{align}
	z_{i,1} = \begin{cases}
		1 & \text{if trial $i$ targets children aged 12-18 years}\\
		0 & \text{if trial $i$ targets children aged 5-11 years}.
	\end{cases}
\end{align}

\textit{Follow-up time covariates:} In the data set, follow-up times are categorized as short (12 weeks to $<$ 9 months), medium (9 months to $<$ 15 months), and long term ($\geq$ 15 months). The follow-up time covariates are therefore defined as
\begin{align}
	(w_{i,1t}, w_{i,2t}) = \begin{cases}
		(0,0) & \text{if follow-up time $t$ is short term}\\
		(1,0) & \text{if follow-up time $t$ is medium term}\\
		(0,1) & \text{if follow-up time $t$ is long term}.
	\end{cases}
\end{align}

\textit{Interactions:} Finally, we choose two interaction terms to investigate whether the effects of diet and activity vary with age. That is,
\begin{align}
	&\mathcal{J}_{i,1t}^{(k)} = x_{i,1}^{(k)}z_{i,1} = \begin{cases}
		1 & \text{if $k$ targets activity only, and $i$ targets children aged 12-18} \\
		0 & \text{otherwise},
	\end{cases} \nonumber \\ 
&\mathcal{J}_{i,2t}^{(k)} = x_{i,2}^{(k)}z_{i,1} = \begin{cases}
	1 & \text{if $k$ targets both diet and activity, and $i$ targets children aged 12-18} \\
	0 & \text{otherwise}.
\end{cases}
\end{align}

Inserting these covariates into Equation (\ref{eq:model-summary}) gives the regression model,
\begin{align}
	\theta_{i,t}^{(k)} = \begin{cases}
		\alpha + \beta_1 x_{i,1}^{(k)} + \beta_2 x_{i,2}^{(k)} + \beta_3 x_{i,3}^{(k)} + \beta_4 x_{i,4}^{(k)} + \gamma_1 z_{i,1} + \phi_1 w_{i,1t} +  \phi_2 w_{i,2t} & \text{} \\ \hspace{185pt} + \eta_1 x_{i,1}^{(k)} z_{i,1}  + \eta_2 x_{i,2}^{(k)} z_{i,1} & \text{for } r=C\\
		\beta_1 (x_{i,1}^{(k)} - x_{i,1}^{(r)})+ \beta_2 (x_{i,2}^{(k)} - x_{i,2}^{(r)}) + \beta_3 (x_{i,3}^{(k)}-x_{i,3}^{(r)}) + \beta_4 (x_{i,4}^{(k)}-x_{i,4}^{(r)}) & \text{} \\ \hspace{120pt} + \eta_1 z_{i,1}(x_{i,1}^{(k)}  - x_{i,1}^{(r)} ) + \eta_2 z_{i,1} (x_{i,2}^{(k)} - x_{i,2}^{(r)}) & \text{for } r=A_i.
	\end{cases}
\end{align}

To specify the covariance matrix $\boldsymbol{V}_i$, we choose correlations $\rho_{y,tt'}$ and $\rho_{d,tt'}$ based on observed correlations in the data set. For time points with one `degree of separation', i.e. short to medium term and medium to long term, we choose $\rho_{y,tt'}=\rho_{d,tt'}=0.8$. Assuming correlations act multiplicatively, we set $\rho_{y,tt'}=\rho_{d,tt'}=0.8^2=0.64$ between short and long term (two degrees of separation). 

\subsubsection{Implementation}
We implement the model in a Bayesian framework using JAGS.\cite{JAGS} In Appendix \ref{App:JAGScode} we provide an example of our JAGS model, defined using the hierarchical structure outlined in Equation (\ref{eq:model-summary}). We assign non-informative prior distributions to all parameters,
\begin{align}
	&\tau \sim \text{Unif}(0,5),\\
	&\alpha, \beta_1, \beta_2, \beta_3, \beta_4, z_1, w_1, w_2, \eta_1, \eta_2 \sim N(0,100^2),
\end{align}
where an upper limit of 5 for heterogeneity and a standard deviation of 100 is large on the zBMI scale which follows a standard normal distribution in the general population. 

We fit the model using Markov chain Monte Carlo (MCMC) sampling. We use an adaptive phase of 10,000 iterations, a burn-in of an additional 10,000, and a further 20,000 iterations from which we draw our posterior samples. We use four chains and assess convergence by inspecting the MCMC trace plots and using the Brookes-Gelman-Rubin $\hat{R}$ statistic.\cite{Brooks:Gelman:1998} To aid convergence we centre all covariates (including interactions) about their mean. This only affects the interpretability of the intercept, $\alpha$, which is not a primary parameter of interest. 

\subsection{Results}\label{sec:Results}
The parameter estimates from the preliminary analysis of the obesity data set are shown in Table \ref{Tab:Results}. We also present the probability that a parameter is less than or greater than zero, calculated from the number of MCMC samples that fall either side of the null. Because our outcomes are a mean difference in change from baseline in zBMI, negative covariate parameters indicate a beneficial effect in terms of obesity prevention. In Appendix \ref{App:Figs} we provide the posterior density and convergence plots for each parameter.

\begin{table}[ht]
	\centering
	\begin{threeparttable}
		\caption{Model parameter estimates for the obesity data set.} 
		\begin{tabular}{l c c c c c} 
			\hline\hline \\ [-1.5ex]
			& Parameter & Median & 95\% CI & $P(<0)$ & $P(>0)$ \\ [0.5ex] 
			\hline\hline \\ [-1.5ex]  
			Intercept (centred) & $\alpha$ & -0.0412 & [-0.0558, -0.0267]& 1.00 & 0.00\\
			Heterogeneity & $\tau$ & \phantom{-}0.0519 & [\phantom{-}0.0409, \phantom{-}0.0645] & 0.00 & 1.00\\
			\multicolumn{6}{l}{\textit{Intervention-level covariates:}}\\
			Activity only & $\beta_1$ & \phantom{-}0.0038 & [-0.0482, \phantom{-}0.0552] & 0.44 & 0.56\\
			Diet \& activity & $\beta_2$ & \phantom{-}0.0105 & [-0.0342, \phantom{-}0.0543] & 0.32 & 0.68\\ 
			Intensity &  $\beta_3$ & -0.0224 & [-0.0510, \phantom{-}0.0058]& 0.94 & 0.06\\
			Duration & $\beta_4$ & \phantom{-}0.0027 & [-0.0311, \phantom{-}0.0368]& 0.44 & 0.56\\
			\multicolumn{6}{l}{\textit{Study-level covariates:}}\\
			Age & $\gamma_1$ & -0.0353 & [-0.1088, \phantom{-}0.0354]& 0.83 & 0.17 \\
			\multicolumn{6}{l}{\textit{Time-level covariates:}}\\
			Medium term & $\phi_1$ & -0.0085 & [-0.0356, \phantom{-}0.0187]& 0.73 & 0.27\\
			Long term & $\phi_2$ & -0.0068 & [-0.0368, \phantom{-}0.0232]& 0.67 & 0.33\\
			\multicolumn{6}{l}{\textit{Interactions:}}\\ 
			Activity only with Age & $\eta_1$ & 0.0889 & [-0.0035, \phantom{-}0.1822]& 0.03 & 0.97\\
			Diet \& activity with Age & $\eta_2$ & 0.0405 & [-0.0451, \phantom{-}0.1270]& 0.18 & 0.82\\
			\hline
		\end{tabular}
        \label{Tab:Results} 
		\begin{tablenotes}
			\footnotesize
			\item CI denotes `credible interval'.
			\item $P(<0)$ and $P(>0)$ represent the probability that the parameter estimate is less than or greater than zero respectively.
		\end{tablenotes}
	\end{threeparttable}
\end{table}

Although none of the covariate parameters show `significant' results in the sense that their credible intervals all cross the null, the posterior densities point to some possible effects. The results suggest little difference in effect between diet only, activity only and combined diet and activity. However, the interaction terms suggest that these results differ according to age. Both the interaction terms are positive, which indicates a reduced benefit in the older age group for the ``activity only" and ``diet and activity" interventions compared with ``diet only". The probability of an effect in this direction is 0.97 for ``activity only" and 0.82 for ``diet and activity". On the other hand, the results for age itself suggests the older age group may benefit more on average than the younger age group.  

While there is no evidence for the effect of duration, the posterior density for ``intensity" indicates that high intensity interventions are more beneficial, $P(\beta_3<0)=0.94$. There is also weak evidence to suggest greatest effects at medium term, followed by long term and then short term.  

In fitting the model, we centred the covariates to improve convergence. Therefore, the intercept can no longer be interpreted as the effect of an intervention whose covariates are all equal to zero. Instead, $\alpha$ represents the effect of covariates at their mean value. This can be thought of as the average effect of interventions. Indeed, the value of $\alpha=-0.04$ is typical of the estimated effects on the zBMI scale we found in our previous analysis of this data set.\cite{Cochrane5to11, Cochrane12to18} Therefore, the results indicate that obesity prevention interventions are effective on average. Finally, there is a strong evidence for non-zero heterogeneity, $\tau$. 

\section{Summary and discussion}\label{sec:Discuss}

We have developed a new network meta-regression-style model for analysing trials of complex interventions. The model facilitates the synthesis of interventions with shared characteristics, allowing us to identify the most effective features of the interventions. By specifying different regression models for trials with and without a control arm, the model relaxes the assumption from previous CNMA models that a control arm is the absence of intervention components. Furthermore, we define a correlation structure to allow for trials with multiple observations at different follow-up times, as well as multiple arms. For maximum flexibility, the model also includes study-level characteristics and interactions between any combination of covariates. 

Using a select subset of covariates, we demonstrated the implementation of our model on data from a systematic review of interventions to prevent obesity in children. We found that interventions are effective on average, in agreement with our previous analysis of this data,\cite{Cochrane5to11, Cochrane12to18} and that high intensity is the strongest indicator of beneficial effect among the covariates investigated. The results for the interaction terms indicate that younger children (aged 5-11 years) may benefit more than older children (aged 12-18 years) from activity interventions. However, we found strong evidence for a non-zero value of $\tau$ meaning there is still unexplained heterogeneity in the model. This suggests that additional covariates are needed to model variation between the trials. In future work, we will perform a more detailed analysis of this data set. This will involve a larger set of covariates and interactions identified using a systematic covariate selection process on the full analytic framework.\cite{AnalysisPlan, Efthimiou:2022} This analysis will be supplemented by additional data from outcomes on percentile and non-standardized BMI scales using novel mapping methods. \cite{AnalysisPlan} In this more detailed analysis we will also investigate the impact of our chosen correlation coefficients.  

Although the model was developed for the specifics of the obesity data set, it has wider applicability to any set of trials comparing complex interventions which can be coded according to a set of shared features. We have assumed binary covariates throughout to aid interpretation and model fitting, but this is not a requirement of the model. 

To allow for multiple observations at different times, we treated follow-up time as a categorical covariate and made the assumption that each trial reports only one observation per time window. The latter assumption can be relaxed by imputing a correlation between observations made in the same time frame (or assuming a correlation of one). An extension to the model might be to treat follow-up time as a continuous variable, allowing for more sophisticated longitudinal modelling. We would then need to choose a function to describe the behaviour of the observations over time. Our results for medium and long term follow-up suggest that such a function should perhaps allow for an initial increase in effectiveness followed by a subsequent decrease, e.g. perhaps a cubic spline with a central knot. However, when using longitudinal models we would have to consider carefully how to define the correlation structure in the presence of multi-arm, multi-follow-up trials. In the obesity data set, most trials report results at only one or two follow-up times. Therefore, we believe there is not much to be gained from more complex longitudinal modelling of these data.

In summary, we have developed and demonstrated the use of a new complex meta-regression model for analysing data from trials of heterogeneous interventions for preventing childhood obesity. The methodology developed goes further than existing models and provides a tool for more detailed analysis of this and other similar data sets.

\section*{Acknowledgments}
ALD and JPTH acknowledge funding from the National Institute for Health and Care Research (NIHR) Public Health Research programme (NIHR131572). JPTH is an NIHR Senior Investigator. We are grateful to Deborah Caldwell for useful discussions.

\section*{Data availability statement}
The JAGS code to implement the model developed in this paper is provided in an appendix. The data and full analysis code that support the findings of this study are available from the corresponding author upon request. 

\section*{Conflicts of Interest}
The authors report no conflict of interests.

\nocite{*}
\bibliography{bibliography}%

\appendix

\section{Random effects covariance matrices}\label{App:RE}
\subsection{Multi-arm, single follow-up trials} \label{App:RE-arm}

In this section we explain how we model correlations between the random trial-specific relative effects, $\delta_{i,t}^{(k)}$, in multi-arm trials. We begin by writing the dependence on the reference arm and baseline time explicitly,
\begin{align}
	\delta_{i,t}^{(k)} = \delta_{i,tb}^{(kr)}.
\end{align}
Recall that $\delta_{i,tb}^{(kr)}$ is a parameter representing the unknown `true' relative effect (mean difference) between arm $k$ and reference arm $r$ in trial $i$. In other words, it is the value that would be measured if $y_{i,tb}^{(kr)}$ had no associated sampling error. We can define similar parameters, $\Delta_{i,tb}^{(k)}$ and  $\mu_{i,t}^{(k)}$, associated with the arm-level observations of change score, $d_{i,tb}^{(k)}$, and time-specific means, $m_{i,t}^{(k)}$, such that
\begin{align}
	\delta_{i,tb}^{(kr)} = \Delta_{i,tb}^{(k)} - \Delta_{i,tb}^{(r)} = (\mu_{i,t}^{(k)}-\mu_{i,b}^{(k)}) - (\mu_{i,t}^{(r)}-\mu_{i,b}^{(r)}). \label{eq:Mdef}
\end{align}
This expression is the parametric equivalent of Equation (\ref{eq:y_deconstruct}) in the main text, relating the observed values.  We will use this deconstruction in Section \ref{App:RE-FU} to demonstrate our assumptions regarding correlations between the $\delta_{i,tb}^{(kr)}$.

The main assumption underlying network meta-analysis, which follows directly from Equation (\ref{eq:Mdef}), is that the trial-specific relative effects are transitive, 
\begin{align}\label{eq:trans-multi-arm}
	\delta_{i,tb}^{(kk')} = \delta_{i,tb}^{(kr)} - \delta_{i,tb}^{(k'r)}.
\end{align}
As mentioned in the main text, we also make the standard assumption from NMA of common between-trial heterogeneity.\cite{Hig:White:1996} That is, we assume that the variation in $\delta_{i,tb}^{(kr)}$ is the same for all interventions $k$ and takes the value $\tau^2$. We further assume that this variation is characteristic of the relative effect between \textit{any pair} of treatments, $\delta_{i,tb}^{(kk')}$. Following from Equation (\ref{eq:trans-multi-arm}), we use the standard variance relation to find
\begin{align}
	\text{var}(\delta_{i,tb}^{(kk')}) &= \text{var}(\delta_{i,tb}^{(kr)}) + \text{var}(\delta_{i,tb}^{(k'r)}) - 2\rho \sqrt{\text{var}(\delta_{i,tb}^{(kr)})\text{var}(\delta_{i,tb}^{(k'r)})}.
\end{align}
Setting $\text{var}(\delta_{i,tb}^{(kk')})=\text{var}(\delta_{i,tb}^{(kr)})=\text{var}(\delta_{i,tb}^{(k'r)})=\tau^2$ gives
\begin{align}
	\tau^2 = 2\tau^2 - 2\rho \tau^2 = 2\tau^2(1-\rho),
\end{align}
which indicates a correlation of $\rho=1/2$. 

In summary,
\begin{align}
	\text{var}(\delta_{i,t}^{(k)}) &= \tau^2, \\
	\text{cov}(\delta_{i,t}^{(k)}, \delta_{i,t}^{(k')}) &= \rho \tau^2 = \tau^2 /2.
\end{align}
Therefore, the covariance matrix $\boldsymbol{\Sigma}_i$ for multi-arm trials has dimensions $(A_i-1)\times(A_i-1)$ with diagonal elements equal to $\tau^2$ and off-diagonal elements equal to $\tau^2/2$.

\subsection{Multi-arm, multi-follow-up trials}\label{App:RE-FU}
To derive correlations between the random effects for multi-follow-up trials we follow the same methodology as in Section \ref{App:RE-arm}. 

\medskip
\noindent\textit{Same intervention at different time points.} Assuming a common intervention $k$, we begin with Equation (\ref{eq:Mdef}) and show that the transitivity relation holds for different follow-up times,
\begin{align}
	\delta_{i,tb}^{(kr)} - \delta_{i,t'b}^{(kr)} &= (\Delta_{i,tb}^{(k)} - \Delta_{i,tb}^{(r)}) - (\Delta_{i,t'b}^{(k)} - \Delta_{i,t'b}^{(r)})\\
	&=((\mu_{i,t}^{(k)}-\mu_{i,b}^{(k)}) - (\mu_{i,t}^{(r)}-\mu_{i,b}^{(r)})) - ((\mu_{i,t'}^{(k)}-\mu_{i,b}^{(k)}) - (\mu_{i,t'}^{(r)}-\mu_{i,b}^{(r)}))\\
	&=(\mu_{i,t}^{(k)}-\mu_{i,t'}^{(k)}) - (\mu_{i,t}^{(r)}-\mu_{i,t'}^{(r)})\\
	&= \Delta_{i,tt'}^{(k)} - \Delta_{i,tt'}^{(r)}\\
	&=\delta_{i,tt'}^{(kr)}.
\end{align}
As before this leads to 
\begin{align}
	\text{var}(\delta_{i,tt'}^{(kr)}) &= \text{var}(\delta_{i,tb}^{(kr)}) + \text{var}(\delta_{i,t'b}^{(kr)}) - 2\rho \sqrt{\text{var}(\delta_{i,tb}^{(kr)})\text{var}(\delta_{i,t'b}^{(kr)})}.
\end{align}
Building on the standard NMA assumption, it is natural to assume that the variation $\text{var}(\delta_{i,tt'}^{(kr)})$ of the relative effect between any pair of time points $(t,t')$ is also equal to $\tau^2$. That is, the between trial variation is the same for all time points $t$. This implies a correlation $\rho=1/2$ for random effects associated with intervention $k$ at different time points. 

\medskip
\noindent\textit{Different interventions at different time points.} For different interventions at different time points we can derive the transitivity relation in two ways, first
\begin{align}
	\delta_{i,tt'}^{(kr)}-\delta_{i,tt'}^{(k'r)} &= (\Delta_{i,tt'}^{(k)}-\Delta_{i,tt'}^{(r)}) -(\Delta_{i,tt'}^{(k')}-\Delta_{i,tt'}^{(r)}) \\
	&=\Delta_{i,tt'}^{(k)}-\Delta_{i,tt'}^{(k')}\\
	&=\delta_{i,tt'}^{(kk')},
\end{align}
or, alternatively, 
\begin{align}
	\delta_{i,tb}^{(kk')}-\delta_{i,t'b}^{(kk')} &= (\Delta_{i,tb}^{(k)}-\Delta_{i,tb}^{(k')}) -(\Delta_{i,t'b}^{(k)}-\Delta_{i,t'b}^{(k')}) \\
	&=((\mu_{i,t}^{(k)} - \mu_{i,b}^{(k)})-(\mu_{i,t}^{(k')} - \mu_{i,b}^{(k')})) - ((\mu_{i,t'}^{(k)} - \mu_{i,b}^{(k)})-(\mu_{i,t'}^{(k')} - \mu_{i,b}^{(k')}))\\
	&=(\mu_{i,t}^{(k)} - \mu_{i,t'}^{(k)}) - (\mu_{i,t}^{(k')}-\mu_{i,t'}^{(k')})\\
	&=\Delta_{i,tt'}^{(k)}-\Delta_{i,tt'}^{(k')}\\
	&=\delta_{i,tt'}^{(kk')}.
\end{align}
From these relations we can define,
\begin{align}
	\text{var}(\delta_{i,tt'}^{(kk')}) &= \text{var}(\delta_{i,tt'}^{(kr)}) + \text{var}(\delta_{i,tt'}^{(k'r)}) - 2\rho \sqrt{\text{var}(\delta_{i,tt'}^{(kr)})\text{var}(\delta_{i,tt'}^{(k'r)})},\\
	\text{var}(\delta_{i,tt'}^{(kk')})&= \text{var}(\delta_{i,tb}^{(kk')}) + \text{var}(\delta_{i,t'b}^{(kk')}) - 2\rho \sqrt{\text{var}(\delta_{i,tb}^{(kk')})\text{var}(\delta_{i,t'b}^{(kk')})}.
\end{align}
Assuming, $\text{var}(\delta_{i,tt'}^{(kk')})=\tau^2$, both relations imply a correlation of $\rho=1/2$ for different interventions at different time points. 

Therefore, the covariance matrix for a trial with multiple arms and multiple follow-up times has dimensions $T_i(A_i-1)\times T_i(A_i-1)$, with diagonal elements equal to $\tau^2$ and off-diagonal elements equal to $\tau^2/2$. 

\section{Within study covariances for multi-arm, multi-follow-up trials} \label{App:Covs}
In this Appendix we derive expressions for the covariances of multi-arm, multi-follow-up trials specified in Equations (\ref{eq:Vit}) and (\ref{eq:Vittprime}). We focus on covariances due to within-study sampling only. For brevity we will suppress the conditioning on the trial-specific relative effects and write
\begin{align}
	\text{cov}(y_{i,t}^{(k)}, y_{i,t'}^{(k')}) \equiv \text{cov}(y_{i,t}^{(k)}, y_{i,t'}^{(k')}|\boldsymbol{\delta}_i).
\end{align}
In the following, we evaluate the covariances for each of the conditions 1-4 in the Section \ref{sec:MultiArm-MultiFU}.

\medskip
1. $t=t'$, $k=k'$: the same time point and same intervention (diagonal elements of $\boldsymbol{V}_{i,t}$). These elements are simply the sampling variance on the measurement in arm $k$ at time $t$, 
\begin{align}
	\text{cov}(y_{i,t}^{(k)}, y_{i,t}^{(k)}) = v_{i,t}^{(k)}.
\end{align}

2. $t=t'$, $k\neq k'$: the same time point and different interventions (off-diagonal elements of $\boldsymbol{V}_{i,t}$). Using the definition of mean difference, $y_{i,t}^{(k)} = d_{i,tb}^{(k)} - d_{i,tb}^{(r)}$, we evaluate
\begin{align}
	\text{cov}(y_{i,t}^{(k)}, y_{i,t'}^{(k')}) &= \text{cov}(d_{i,tb}^{(k)}- d_{i,tb}^{(r)}, d_{i,t'b}^{(k')}- d_{i,t'b}^{(r)})\nonumber\\
	&= \text{cov}( d_{i,tb}^{(k)} , d_{i,t'b}^{(k')} ) - \text{cov}( d_{i,tb}^{(k)} , d_{i,t'b}^{(r)} ) - \text{cov}( d_{i,tb}^{(r)} , d_{i,t'b}^{(k')} ) + \text{cov}( d_{i,tb}^{(r)} , d_{i,t'b}^{(r)} )\nonumber\\
	&= I(k=k') \text{cov}( d_{i,tb}^{(k)} , d_{i,t'b}^{(k')} )  + \text{cov}( d_{i,tb}^{(r)} , d_{i,t'b}^{(r)} ), \label{eq:cov}
\end{align}
where we have used the fact that change scores in different arms are independent. We have also made use of the indicator, $I(.)$, defined as $I(A)=1$ when $A$ is true and $I(A)=0$ when $A$ is false. Setting $t=t'$ and $k\neq k'$ we find 
\begin{align}
	\text{cov}(y_{i,t}^{(k)}, y_{i,t}^{(k')}) &= \text{cov}( d_{i,tb}^{(r)} , d_{i,tb}^{(r)} ) =\text{var}(d_{i,tb}^{(r)}),
\end{align}
which is the variance of the change score in the reference arm at time $t$. As expected, this is the same as the off-diagonal elements of the covariance matrix in Section \ref{sec:MultiArm-SingleFU} of the main text.

3. $t\neq t'$, $k = k'$: different time points and the same intervention (diagonal elements of $\boldsymbol{V}_{i,tt'}$). Using the definition of covariance we find, 
\begin{align}
	\text{cov}(y_{i,t}^{(k)}, y_{i,t'}^{(k)}) &= \rho_{y,tt'} \sqrt{\text{var}(y_{i,t}^{(k)})\text{var}(y_{i,t'}^{(k)})} = \rho_{y,tt'} \sqrt{v_{i,t}^{(k)}v_{i,t'}^{(k)}},
\end{align}
where $\rho_{y,tt'}$ is the correlation between observations of mean difference for a given contrast (arm $k$ vs the reference arm $r$) at time points $t$ and $t'$. We have assumed that these correlations are the same for all arms, i.e. $\rho_{y,tt'}^{(k)} = \rho_{y,tt'} \forall k$. In general, the correlation $\rho_{y,tt'}$ is unknown but can be estimated from the data.

4. $t\neq t'$, $k\neq k'$: different time points and different interventions (off-diagonal elements of $\boldsymbol{V}_{i,tt'}$). Specifying $t\neq t'$ and $k\neq k'$ in Equation (\ref{eq:cov}) we find
\begin{align}
	\text{cov}(y_{it}^{(k)}, y_{it'}^{(k')}) &= \text{cov}( d_{i,tb}^{(r)} , d_{i,t'b}^{(r)} ) \\
	&=\rho_{d,tt'} \sqrt{\text{var}(d_{i,tb}^{(r)})\text{var}(d_{i,t'b}^{(r)})},
\end{align}
where $\rho_{d,tt'}$ is the correlation between observations of the change score in the reference arm at time points $t$ and $t'$. Where arm-level observations are available, this correlation can be imputed from the data.

\section{Covariance matrices in active comparison studies}\label{App:Cov_active}
\subsection{Random effects covariance matrix}
To derive the covariance matrix $\boldsymbol{\Sigma}_i$ in Equation (\ref{eq:active-full}) for active comparison studies, we start from the transitivity relation in Equation (\ref{eq:delta_trans}). Using the standard rules for combining variances and covariances we find
\begin{align}
	\text{var}(\delta_{i,t}^{(k)}-\delta_{i,t}^{(A_i)}) &= \text{var}(\delta_{i,t}^{(k)}) + \text{var}(\delta_{i,t}^{(A_i)}) - 2\text{cov}(\delta_{i,t}^{(k)},\delta_{i,t}^{(A_i)}) \\
	&= \tau^2 + \tau^2 -2 \frac{\tau^2}{2} = \tau^2,
\end{align}
and 
\begin{align}
	\text{cov}(\delta_{i,t}^{(k)}-\delta_{i,t}^{(A_i)}, \delta_{i,t}^{(k')}-\delta_{i,t}^{(A_i)}) 
	&= \text{cov}(\delta_{i,t}^{(k)}, \delta_{i,t}^{(k')}) - \text{cov}(\delta_{i,t}^{(k)}, \delta_{i,t}^{(A_i)}) - \text{cov}(\delta_{i,t}^{(A_i)}, \delta_{it}^{(k')}) + \text{cov}(\delta_{i,t}^{(A_i)},\delta_{i,t}^{(A_i)})\\
	&=\frac{\tau^2}{2} - \frac{\tau^2}{2} - \frac{\tau^2}{2} + \tau^2 = \frac{\tau^2}{2}.
\end{align}
Therefore, the $T_i(A_i-1)$-vector $\boldsymbol{\delta}_i$ in an active comparison trial follows a normal distribution with variances equal to $\tau^2$ and covariances equal to $\tau^2/2$. The covariance matrix $\boldsymbol{\Sigma}_i$ is therefore the same as in the control comparison model (Equation (\ref{eq:RE_multi1})). 

\subsection{Sampling error covariance matrix}
The covariance matrix $\boldsymbol{V}_i$ characterizes the within-trial sampling variances and correlations. For active comparison studies, the diagonal elements are simply the variances on the outcome measurements, $v_{i,t}^{(k)} = v_{i,tb}^{(kA_i)} = \text{var}(y_{i,tb}^{(kA_i)})$. The off-diagonal elements are the covariances, $\text{cov}(y_{i,tb}^{(kA_i)}, y_{i,t'b}^{(k'A_i)})$. Using Equation (\ref{eq:y-trans}) in the main text and standard covariance rules we find
\begin{align}
	\text{cov}(y_{i,tb}^{(kA_i)}, y_{i,t'b}^{(k'A_i)}) &= \text{cov}(y_{i,tb}^{(kC)} - y_{i,tb}^{(A_iC)} , y_{it'b}^{(k'C)} - y_{it'b}^{(A_iC)})\nonumber\\
	&= \text{cov}((d_{i,tb}^{(k)}- d_{i,tb}^{(C)}) - (d_{i,tb}^{(A_i)}- d_{i,tb}^{(C)}), (d_{i,t'b}^{(k')}- d_{i,t'b}^{(C)}) - (d_{i,t'b}^{(A_i)}- d_{i,t'b}^{(C)}) )\nonumber\\
	&=\text{cov}(d_{i,tb}^{(k)} - d_{i,tb}^{(A_i)}, d_{i,t'b}^{(k')} - d_{i,t'b}^{(A_i)}) \nonumber\\
	&= \text{cov}( d_{i,tb}^{(k)} , d_{i,t'b}^{(k')} ) - \text{cov}( d_{i,tb}^{(k)} , d_{i,t'b}^{(A_i)} ) - \text{cov}( d_{i,tb}^{(A_i)} , d_{i,t'b}^{(k')} ) + \text{cov}( d_{i,tb}^{(A_i)} , d_{i,t'b}^{(A_i)} ) \nonumber\\
	&= I(k=k') \text{cov}( d_{i,tb}^{(k)} , d_{i,t'b}^{(k')} )  + \text{cov}( d_{i,tb}^{(A_i)} , d_{i,t'b}^{(A_i)} ). 
\end{align}
This is the same expression we derived in Appendix \ref{App:Covs} for the control comparison case (Equation (\ref{eq:cov})) but with $r=A_i$. Therefore, we can  model the within-study correlations in the same way in all studies.

\section{JAGS code}\label{App:JAGScode}
An example of the JAGS model used to analyse the obesity data set with the parameters described in Section \ref{sec:Methods}. 

\begin{lstlisting}
model{
  for(i in 1:N){
    #dim[i] is the dimensions of trial i = Ti(Ai-1)
    #y and V are ragged arrays containing the observations of MD and the within study covariance matrix
  
    # multivariate normal: yi ~ N(deltai, Vi)
    y[yind[i]:yend[i]]~dmnorm.vcov(delta[yind[i]:yend[i]], V[i,1:dim[i],1:dim[i]])
	
    # define the RE covariance matrix in terms of tau and S
    # S is a ragged array containing matrices with diag = 1, off-diag = 0.5
    Sigma[i,1:dim[i],1:dim[i]]<-(tau^2)*S[i,1:dim[i],1:dim[i]]
	
    # multivariate normal (RE): deltai ~ N(thetai, Sigmai)
    delta[yind[i]:yend[i]]~dmnorm.vcov(theta[yind[i]:yend[i]], Sigma[i,1:dim[i],1:dim[i]])
	
    # Now define theta (regression model)
    # numFU[i] is the number of follow-up times in trial i
    # numcont[i] is the number of contrasts (arms - 1) in trial i
    for(t in 1:numFU[i]){ 
      for(k in 1:numcont[i]){        
        # terms for both active and control comparisons (x and J)----------
        #NB: for active comparison trials, x and J are defined as ...
        #...the difference between arm k and the reference arm 
        
        #x (interventions)
        sum_x[i,t,k] <- inprod(beta, x[i,k, ])
			
        #J (interactions)
        sum_J[i,t,k] <- inprod(eta, J[i,t,k, ])
			
        # control comparison only terms (alpha, z, w)--------------
			
        #z (study-level)
        sum_z[i,t,k] <- inprod(gamma, z[i, ])
			
        #w (FU time)
        sum_w[i,t,k] <- inprod(phi, w[i,t, ])
			
        #ref_arm is an indicator: control comparison=1, active comparison=0
        #therefore active models only have the first two terms
        theta[tind[i,t,k]] <- sum_x[i,t,k] + sum_J[i,t,k] +
                              ref_arm[i]*(alpha + sum_z[i,t,k] + sum_w[i,t,k])	
      }#end sum over k
    }#end sum over t
  }#end sum over i
	
  ### Define the priors----------------------------------------------------
  #Heterogeneity: tau (uniformative)
  tau ~ dunif(0, 5) 

  #Intercept: alpha ~ N(0, 100^2) (uniformative)
  alpha ~ dnorm(0, 0.0001) 

  #Interventions: beta ~ N(0, 100^2) (uniformative)
  for(j in 1:n){ beta[j] ~ dnorm(0, 0.0001) } 

  #Study covariates: gamma ~ N(0, 100^2) (uniformative)
  for(j in 1:p){ gamma[j] ~ dnorm(0, 0.0001) } 

  #follow-up time: phi ~ N(0, 100^2) (uniformative)
  for(j in 1:(q-1)){ phi[j] ~ dnorm(0, 0.0001) } 

  #Interactions: eta ~ N(0, 100^2) (uniformative)
  for(j in 1:l){ eta[j] ~ dnorm(0, 0.0001) } 
}
\end{lstlisting}

\section{MCMC plots}\label{App:Figs}
\begin{figure}[h]
	\centering
	\includegraphics[width=1\linewidth]{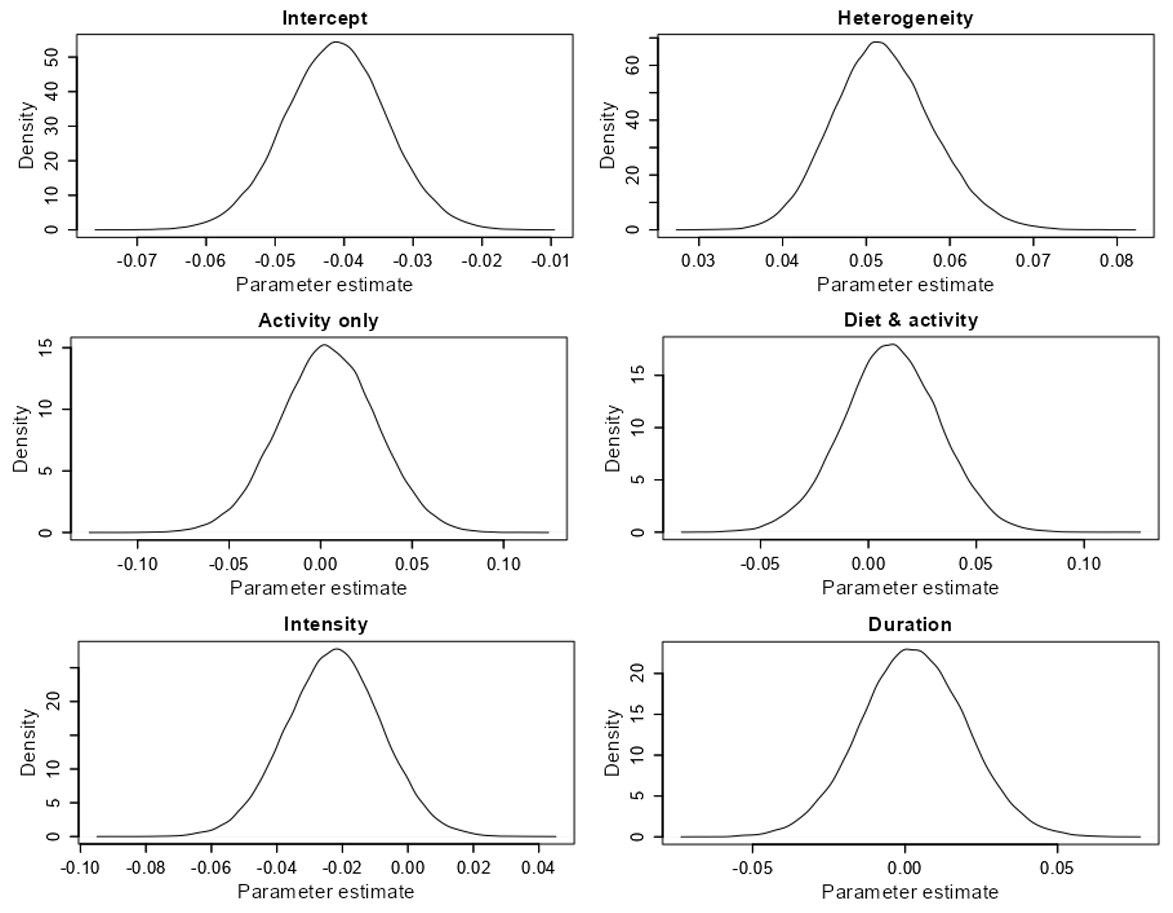}
	\caption{Posterior density plots for parameters $\alpha$ (Intercept), $\tau$ (Heterogeneity), $\beta_1$ (Activity only), $\beta_2$ (Diet \& activity), $\beta_3$ (Intensity) and $\beta_4$ (Duration) in the model fitted to the obesity data set. Parameter estimates based on these posteriors are shown in Table \ref{Tab:Results}.}
	\label{Fig:Density1}
\end{figure}

\begin{figure}[h]
	\centering
	\includegraphics[width=1\linewidth]{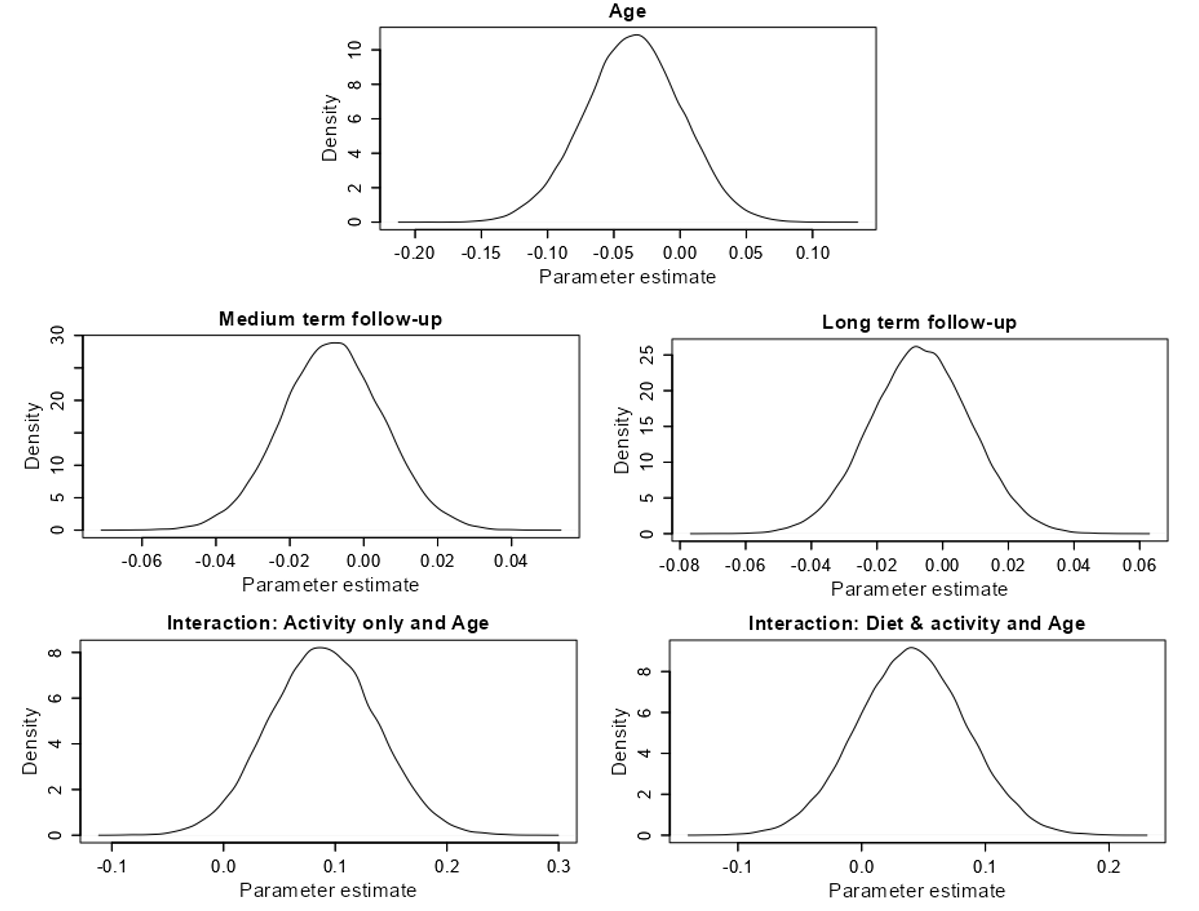}
	\caption{Posterior density plots for parameters $\gamma_1$ (Age), $\phi_1$ (Medium term), $\phi_2$ (Long term), $\eta_1$ (Interaction: Activity only and Age), $\eta_2$ (Interaction: Diet \& activity and Age) in the model fitted to the obesity data set. Parameter estimates based on these posteriors are shown in Table \ref{Tab:Results}.}
	\label{Fig:Density2}
\end{figure}

\begin{figure}[h]
	\centering
	\includegraphics[width=1\linewidth]{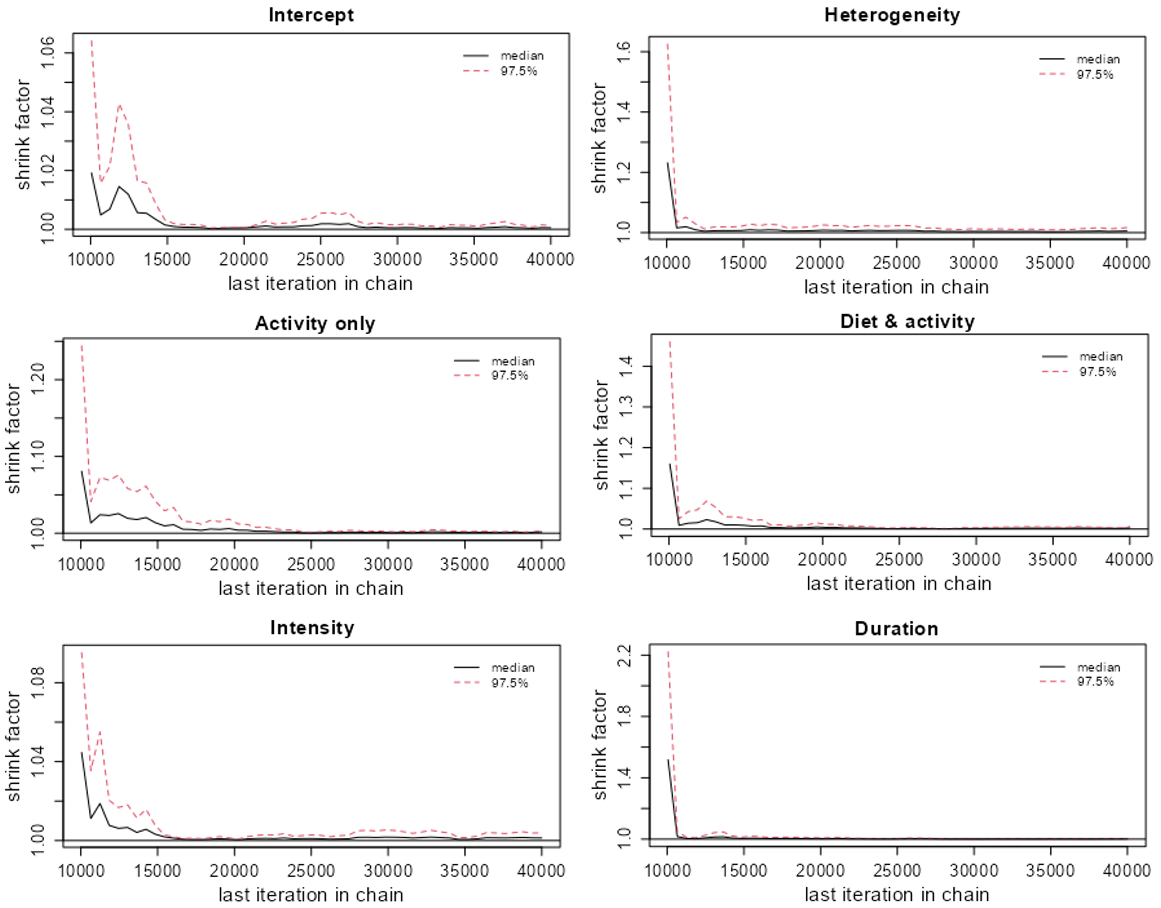}
	\caption{Brooks-Gelman-Rubin convergence plots for parameters $\alpha$ (Intercept), $\tau$ (Heterogeneity), $\beta_1$ (Activity only), $\beta_2$ (Diet \& activity), $\beta_3$ (Intensity) and $\beta_4$ (Duration) in the model fitted to the obesity data set. }
	\label{Fig:Gelman1}
\end{figure}

\begin{figure}[h]
	\centering
	\includegraphics[width=1\linewidth]{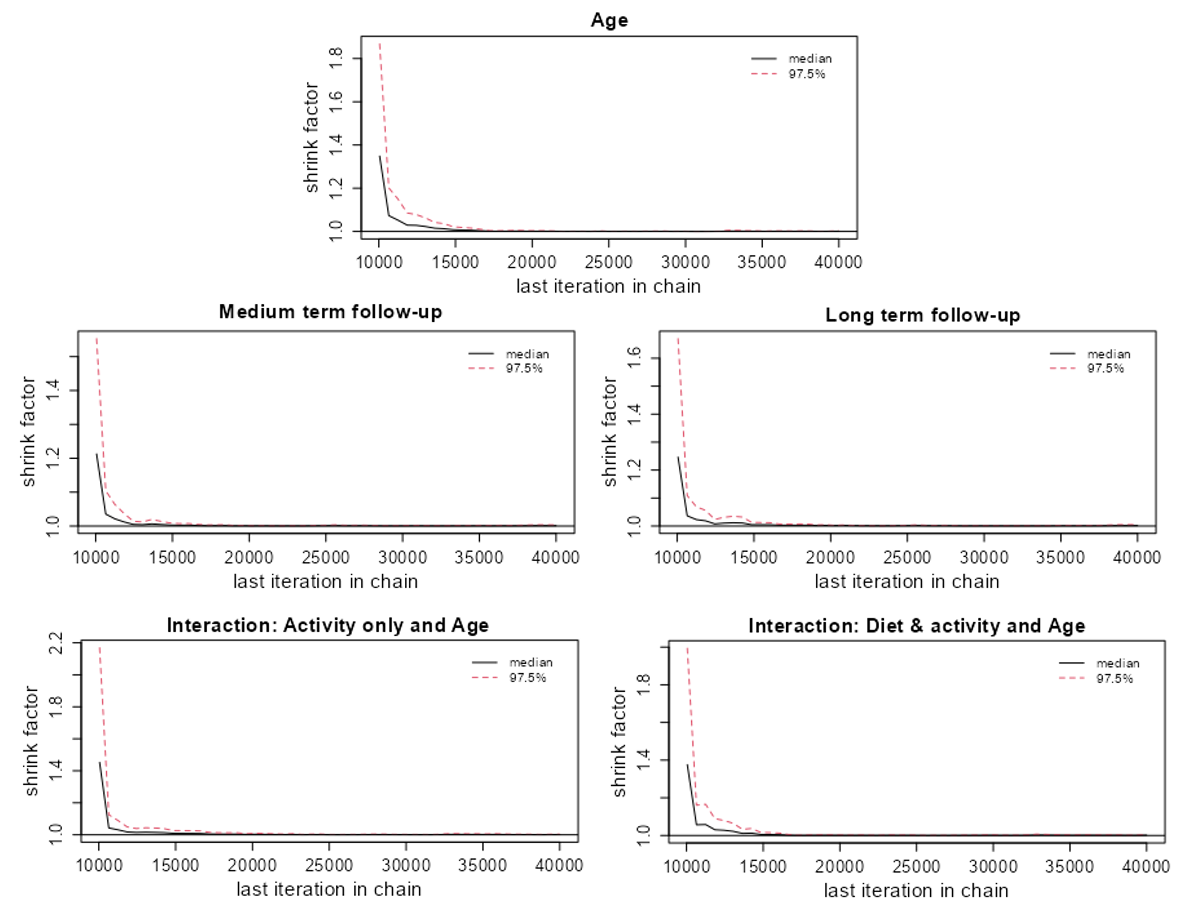}
	\caption{Brooks-Gelman-Rubin convergence plots for parameters $\gamma_1$ (Age), $\phi_1$ (Medium term), $\phi_2$ (Long term), $\eta_1$ (Interaction: Activity only and Age), $\eta_2$ (Interaction: Diet \& activity and Age) in the model fitted to the obesity data set.}
	\label{Fig:Gelman2}
\end{figure}

Figures \ref{Fig:Density1} and \ref{Fig:Density2} show the posterior density plots for the parameters in Table \ref{Tab:Results}. Figures \ref{Fig:Gelman1} and \ref{Fig:Gelman2} show the corresponding Brooks-Gelman-Rubin convergence plots  \cite{Brooks:Gelman:1998} for these parameters. Here we plot the shrink factor\cite{Gelman:Rubin:1992} (ratio of pooled variance and average within-chain variance)  against the number of iterations. Convergence is indicated when the shrink factor stops fluctuating and reaches a value less than approximately 1.1. \cite{Brooks:Gelman:1998, Gelman:Rubin:1992} For all parameters this happens by 20,000 iterations or before. Therefore, we chose a burn-in of 10,000 on top of the adaptive phase of 10,000. 

\end{document}